\documentclass{JHEP3}
\usepackage{amsmath,amssymb}
\usepackage{graphicx}

\newcommand{\BK}{\text{K}}
\newcommand{\BH}{\text{H}}
\newcommand{\BI}{\text{I}}
\newcommand{\BJ}{\text{J}}

\title{Particle production in DIS off a shockwave in AdS}
\author{Fabio Dominguez\\ Department of Physics,
Columbia University, New York, NY, 10027, USA\\ Email: \email{fabio@phys.columbia.edu}}

\abstract{Within the AdS/CFT correspondence, the shockwave metric has been widely used as a gravity dual for a fast moving nucleus. Here we propose a picture for particle production in deep inelastic scattering off a shockwave with the projectile represented by the $\mathcal{R}$-current. By using the method developed in \cite{Avsar:2009xf} to find an explicit expression for the scattered field, we are able to show that the scattered field can be written in terms of time-like and space-like vacuum states where the time-like states are identified with the outgoing particles. To support this picture, we calculate the contribution of the time-like modes to the energy-momentum tensor of the scattered field and show that the energy flow in the fifth dimension is directly related to the imaginary part of the action. The energy flow in the fifth dimension is also compared to the incoming energy flow to determine which fraction of the initial energy goes to the produced particles. The last two sections are devoted to localize the energy flow in momentum and coordinate space, first by finding the virtualities that contribute the most to the total flow, and second by finding an approximate trajectory in coordinate space.}

\begin{document}

\section{Introduction}
In recent years there have been numerous attempts to explain the results of heavy ion collisions suggesting the existence of a strongly interacting deconfined QCD plasma. The complications of dealing with a strongly coupled theory in the framework of familiar quantum field theories have led to the use of methods from string theory via the AdS/CFT correspondence \cite{Maldacena:1997re,Gubser:1998bc,Witten:1998qj,Son:2007vk}. Most of the calculations performed up to date are set in the supergravity scenario where the t'Hooft coupling $\lambda=g^{2}N$ of the gauge theory is assumed to be large by taking the number of colors to infinity while the gauge coupling $g$ is still small. This scenario is chosen for practical purposes, since in that approximation the string theory side of the correspondence is reduced to a supergravity theory (string excitations are suppressed).

The conformally symmetric $\mathcal{N}=4$ SYM is not expected to reproduce accurately all regimes of QCD, but for the case of the strongly coupled plasma created in heavy ion collisions it is widely believed to give sensible results. To be able to use the possible similarities between the two different theories within this context, it is important to find a suitable gravity dual for a fast moving nucleus. A natural choice would be a highly boosted slab of matter, as the one described in \cite{Mueller:2008bt}, but this scenario has the inconvenience of using a metric which is not an exact solution of Einstein's equations. Taking into account that in the limit of an infinite boost the finite slab of matter is supposed to look like a shockwave, it is more convenient to work with a shockwave metric (proved to be an exact solution to the equations of motion). This choice was first suggested in \cite{Janik:2005zt} but since then different alternatives to define more general scenarios have been developed \cite{Beuf:2009mk,Avsar:2009xf,Albacete:2008ze,Gubser:2008pc}.

Different attempts to explore the properties of the shockwave metric in AdS$_{5}$ have led to interesting results. The first computation of the scattering of a field in a shock wave in AdS was done in \cite{Cornalba:2006xk,Cornalba:2006xm}. Subsequent studies in the resummation graviton exchanges and computation of structure functions include \cite{Cornalba:2007zb,Brower:2007qh,Albacete:2008ze,Avsar:2009xf}. In particular we focus on the deep inelastic scattering analysis presented in \cite{Avsar:2009xf} where explicit expressions for the fields representing the probes are calculated but the main focus is in the calculation of the structure functions, which are obtained relying on the optical theorem and using a forward scattering amplitude. In this paper we choose to take a different path, by using the $\mathcal{R}$-current as a probe we follow the propagation of the gravity wave after the scattering process. Using the explicit expressions of the fields calculated in \cite{Avsar:2009xf}, we calculate the energy-momentum tensor associated with the produced states after the collision and give a physical picture of the results.

The original projectile considered is assumed to be a space-like $\mathcal{R}$-current. The scattered field is calculated from a multiple scattering approach which takes the form of an eikonal phase when the shockwave is considered to be a $\delta$-function in $x^{-}$. From the point of view of  the shockwave as a finite slab of matter under a large boost, the eikonal picture is consistent only if the coherence length of the incoming probe is larger than the width of the target. This condition was observed in \cite{Mueller:2008bt}, it also agrees with the analysis of the validity of the $\delta$-function approximation in \cite{Avsar:2009xf}, and it is consistent with what we find in Eq. (\ref{cohlength}) in our attempt to localize the energy flow. When we look at the scattered field we notice it can be regarded as a combination of space-like and time-like vacuum modes. The appearance of the time-like modes after the scattering resembles the known process of particle production in deep inelastic scattering, where real (time-like) particles appear in the final state after a collision involving a highly space-like probe. Based on this observation, we follow the propagation of these produced time-like states and argue that the picture of particle production is applicable in this context.

Given that the calculations are performed in the classical approximation, by solving the classical equations of motion, the particle production picture is not completely clear in the usual sense. Previous calculations regarding scattering processes with various targets \cite{Cornalba:2007zb,Brower:2007qh,Polchinski:2002jw,Levin:2009vj,Hatta:2007cs,'tHooft:1987rb} show that the main contribution to the scattering amplitudes, in the strong coupling limit, come from elastic or quasi-elastic processes where particle production doesn't play an important role since at high energy they are dominated by graviton exchange. Other mechanisms for multiparticle production have been suggested \cite{Kharzeev:2009pa} as an attempt to get a consistent picture for the creation of a quark-gluon plasma. We argue this is not necessary to get a consistent picture of particle production in deep inelastic scattering. In this classical calculation, particle production is manifest through the presence of time-like modes in the expansion of the fields after the scattering.

\begin{figure}
\begin{center}
\includegraphics{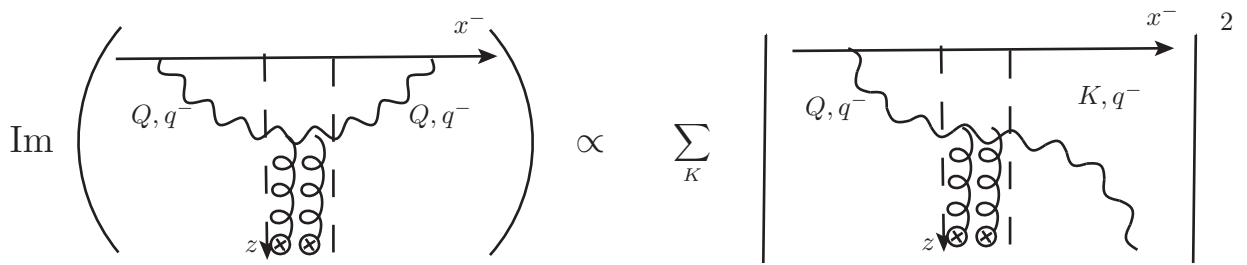}
\end{center}
\caption{The photon line attached to the boundary represents a space-like state, like the incoming wave. The photon line going down the fifth dimension represents a time-like state. The region in between the dashed lines is the shockwave where the graviton exchanges take place.}\label{figot}
\end{figure}

Our argument to support this particle production picture is based on the calculation of the energy-momentum tensor of the $\mathcal{R}$-current. The inelastic part of the scattering, which is associated to the imaginary part of the action, is directly related to the energy flow in the fifth dimension, which is caused by the presence of the time-like modes in the scattered field. As a comparison with previous calculations of the structure functions, this can be seen as an analog of the optical theorem. Our calculation gives the sum over final states of the amplitude squared, which has to be proportional to the imaginary part of the forward current-current correlator (see Fig. \ref{figot}). The energy flow in the fifth dimension is also compared to the initial state where the incoming probe is assumed to be carrying energy in the $x^{-}$ direction. This comparison gives a clear picture of the scattering process in the saturation regime and shows a clear difference in the behavior of the transverse and longitudinal components.

Keeping in mind the picture of the energy flow in the fifth dimension representing the inelastic part of the collision and, therefore, the produced particles, we turn our attention to determine the momentum of the outgoing particles. The explicit expressions for the scattered fields written in terms of vacuum states suggest that the coefficient functions should be considered as the probabilities of the different modes to be produced on any given scattering process. This picture is not completely clear since the basis of vacuum states is not a complete set of generators. To lift this ambiguity we rely on our calculation of the energy-momentum tensor to find which states (labeled by their virtualities) are the ones contributing the most to the energy flow. Given the separation between transverse and longitudinal components on the calculation of the energy-momentum tensor we can find different regions for both cases and find a direct relation with the transverse and longitudinal structure functions found in \cite{Mueller:2008bt,Avsar:2009xf}.

The last section is devoted to find an approximate trajectory of the energy flow in the fifth dimension. Since the incoming probe is assumed to be moving along the $x^{-}$ direction we attempt to give an estimate of where in the $x^{-}$ coordinate is the energy flow localized for any given value of $z$. For large values of $z$ we find a linear relation between $x^{-}$ and $z$ that doesn't depend on the properties of the target. It is also interesting to see how the coherence length of the projectile shows up as a restriction on where the different modes can be resolved, supporting the fact that the eikonal approximation is valid only as long as this coherence length is larger than the size of the target.

\section{General setup and scattered wave}

The general form of the shock wave metric, in Fefferman-Graham coordinates, is \cite{Janik:2005zt}
\begin{equation}
ds^{2}=\frac{R^{2}}{z^{2}}\left(dz^{2}-2dx^{+}dx^{-}+dx_{\perp}^{2}+h(z,x^{-},x_{\perp})(dx^{-})^{2}\right).\label{metric}
\end{equation}
There are different procedures used in the literature \cite{Avsar:2009xf,Beuf:2009mk,Janik:2005zt,Albacete:2008ze,Gubser:2008pc} to determine and physically motivate the appropriate shape of the shock wave, which is encoded in the function $h(z,x^{-},x_{\perp})$. Here, for simplicity, we assume no dependence in the transverse coordinates (an homogeneous infinite wall), in which case the function $h$ takes the simple form
\begin{equation}
h(z,x^{-})=\frac{2\pi^{2}}{N^{2}}z^{4}\langle T_{--}(x^{-})\rangle\, ,\label{homsw}
\end{equation}
where $T_{--}$ on the right hand side corresponds to the gauge-theory energy-momentum tensor which will represent a fast moving nucleus. The $z^{4}$ factor is required for the metric to satisfy Einstein's equation and the relation with the energy-momentum tensor is given by holographic renormalization \cite{de Haro:2000xn,Skenderis:2002wp}.

This form of the metric corresponds to the target infinite momentum reference frame with the target moving in the positive $x^{3}$ direction (right mover). The projectile to be considered is an $\mathcal{R}$-current, represented by a gauge vector field in AdS$_{5}$ with space-like momentum $q^{\mu}$, moving in the negative $x^{3}$ direction (left mover). Taking $q_{\perp}=0$, we have $q^{-}>0$, $q^{+}<0$, and the virtuality $Q^{2}=-2q^{+}q^{-}>0$.

We are interested in the explicit form of the classical field after the scattering. This can be found by using a multiple scattering approach to solve the equations of motion \cite{Avsar:2009xf}, which are derived from the 5-dimensional Yang-Mills action
\begin{equation}
S=-\frac{N^{2}}{64\pi^{2}R}\int d^{4}x\,dz\;\sqrt{-g}g^{mp}g^{nq}F_{mn}F_{pq}\, .\label{action}
\end{equation}
Taking into account that the shockwave metric in Eq. (\ref{metric}) does not depend on $x^{+}$, the $q^{-}$ component of the initial momentum of the projectile will be conserved and, therefore, we can restrict our analysis to fields of the form
\begin{equation}
A_{\mu}(x^{-},x^{+},z)=e^{-iq^{-}x^{+}}A_{\mu}(x^{-},z)\, .\label{planeA}
\end{equation}

For convenience we use the gauge condition $A_{z}=0$. Under these circumstances, the relevant equations of motion are \cite{Avsar:2009xf}
\begin{eqnarray}
(z\partial_{z}z^{-1}\partial_{z}+iq^{-}\partial_{-})A_{+}=(q^{-})^{2}A_{-}\, ,\label{leom1} \\
(\partial_{-}-iq^{-}h)A'_{+}=iq^{-}A'_{-}\, ,\label{leom2} \\
(z\partial_{z}z^{-1}\partial_{z}+2iq^{-}\partial_{-})A_{i}=-(q^{-})^{2}hA_{i}\, ,\label{treom}
\end{eqnarray}
where a prime denotes differentiation with respect to $z$.

Let's focus on the longitudinal components first. Eqs. \eqref{leom1} and \eqref{leom2} can be combined into a single differential equation for $A'_{+}$,
\begin{equation}
(\partial_{z}z\partial_{z}z^{-1}+2iq^{-}\partial_{-})A'_{+}=-(q^{-})^{2}hA'_{+}\, .\label{comblongeom}
\end{equation}
This is equivalent to the integral equation
\begin{equation}
A'_{+}(x^{-},z)=A_{+}^{\prime(0)}(x^{-},z)+\int\frac{dz'}{z'}dy^{-}G_{L}(z,z';x^{-}-y^{-})\left[-(q^{-})^{2}\right]h(z',y^{-})A'_{+}(y^{-},z')\, ,\label{intlong}
\end{equation}
where $A_{+}^{\prime(0)}$ is the vacuum solution, and the Green's function $G_{L}$ satisfies the equation
\begin{equation}
(\partial_{z}z\partial_{z}z^{-1}+2iq^{-}\partial_{-})G_{L}(z,z';x^{-}-y^{-})=z\delta(z-z')\delta(x^{-}-y^{-})\, .\label{greeneq}
\end{equation}

In Eq. (\ref{intlong}) the field is written as a vacuum piece and a scattering piece. Following Ref. \cite{Avsar:2009xf} we choose to impose the boundary condition at $z=0$ (plane wave state) on the vacuum solution, or equivalently, require the Green's function to vanish at the boundary. More specifically, we write the field as
\begin{equation}
A_{\mu}(x^{-},z)=A_{\mu}^{(0)}(x^{-},z)+A_{\mu}^{(s)}(x^{-},z)\, ,
\end{equation}
and the boundary condition reads
\begin{align}
\lim_{z\to0}A_{\mu}^{(0)}(x^{-},z)&=\mathcal{A}_{\mu}(x^{-})\, ,\\
\lim_{z\to0}A_{\mu}^{(s)}(x^{-},z)&=0\, .
\end{align}
The boundary field is assumed to be a plane wave and therefore can be written as
\begin{equation}
\mathcal{A}_{\mu}(x^{-})=e^{-iq^{+}x^{-}}\tilde{\mathcal{A}}_{\mu}\, ,
\end{equation}
with $\tilde{\mathcal{A}}_{\mu}$ pure numbers.

These boundary conditions together with Eq. (\ref{leom1}) imply the following boundary condition for $A'_{+}$
\begin{equation}
\lim_{z\to0}z\partial_{z}(z^{-1}A'_{+}(x^{-},z))=\frac{Q^{2}}{2}\mathcal{A}_{+}(x^{-})+(q^{-})^{2}\mathcal{A}_{-}(x^{-})\equiv \frac{Q^{2}}{2}\mathcal{A}_{L}(x^{-})\, .\label{boundarylong}
\end{equation}

The longitudinal vacuum solution is obtained from the homogeneous version of Eq. (\ref{comblongeom}) and the boundary condition (\ref{boundarylong}) \cite{Hatta:2007cs}
\begin{equation}
A^{\prime(0)}_{+}(x^{-},z)=-\frac{1}{2}Q^{2}\mathcal{A}_{L}(x^{-})z\BK_{0}(Qz)\, .\label{vacsol}
\end{equation}

Here we made use of the fact that the initial momentum of the incoming wave is assumed to be space-like by taking the solution with a K-function (for time-like momentum the appropriate solution involves a Hankel function instead). The distinction between time-like and space-like vacuum solutions will play an important role in our discussion because of their different behavior at large $z$. As will be seen below, the scattering piece of the field can be written in terms of a superposition of time-like and space-like vacuum states, but by taking the large $z$ limit we are able to isolate the contribution from time-like modes only since the space-like ones go exponentially to zero in that region.

Now lets turn our attention to the scattering piece of the field which will be obtained by means of a multiple scattering approach. In particular we will consider the simpler case where the shockwave is represented by a delta function in the $x^{-}$ coordinate, namely $h(x^{-},z)=\delta(x^{-})\tilde{h}(z)$. By considering multiple iterations of Eq. (\ref{intlong}) and taking into account that the Green's function is retarded in the $x^{-}$ variable and satisfies \cite{Avsar:2009xf}
\begin{equation}
G_{L}(z,z';x^{-}-y^{-}\to+0)=-\frac{i}{2q^{-}}z\delta(z-z')\, ,\label{greenzeroright}
\end{equation}
we can construct a series that exponentiates into an eikonal phase. This procedure gives the result
\begin{equation}
A'_{+}(x^{-},z)=A_{+}^{\prime(0)}(x^{-},z)-2q^{-}\int \frac{dz'}{z'}G_{L}(z,z';x^{-})\mathcal{T}(z')A_{+}^{\prime(0)}(0,z')\, ,\label{fullsollg}
\end{equation}
where $\mathcal{T}(z')$ is the scattering amplitude defined by
\begin{equation}
-i\mathcal{T}(z)=1-\exp\left(\frac{iq^{-}}{2}\tilde{h}(z)\right).\label{scatamp}
\end{equation}

Evaluating (\ref{fullsollg}) at $x^{-}=+0$ and using (\ref{greenzeroright}), we can easily see that the fields pick up a phase when going through the shockwave
\begin{equation}
A'_{+}(+0,z)=\exp\left(\frac{iq^{-}}{2}\tilde{h}(z)\right)A'_{+}(-0,z)\, .\label{discphase}
\end{equation}
When this is inserted in Eq. (\ref{planeA}), it can be seen as a jump in the light-cone time $x^{+}$ which corresponds to the capture time already observed in Refs. \cite{Mueller:2008bt,Kancheli:2002nw}.

Using Eq. (\ref{homsw}) and the fact that we are using the shockwave as a gravity dual for a fast moving nucleus, a sensible choice for $\tilde{h}$ would be
\begin{equation}
\tilde{h}(z)=2\pi^{2}z^{4}\gamma L\Lambda^{4}\, ,
\end{equation}
where $\gamma$ is the boost factor, $L$ the size of the target on its rest frame, and $\Lambda$ the characteristic momentum scale in the target rest frame. Introducing the corresponding Bjorken-$x$ variable
\begin{equation}
x=\frac{Q^{2}}{2q^{-}\gamma\Lambda}\, ,
\end{equation}
the exponent in (\ref{scatamp}) can be written as
\begin{equation}
\frac{iq^{-}}{2}\tilde{h}(z)=iQ^{2}\frac{\pi^{2}L\Lambda^{3}}{2x}z^{4}\, .
\end{equation}

The values of $z$ entering the scattering amplitude are typically of order $1/Q$, and therefore the momentum scale which determines when multiple scattering become important is $Q_{s}^2=\pi^{2}L\Lambda^{3}/2x$. From now on we will write the scattering amplitude as\footnote{Here we are following closely the derivation in \cite{Avsar:2009xf}. Saturation effects in an AdS/CFT context were first considered in \cite{Hatta:2007he,Cornalba:2008sp} and subsequently in a more general context in \cite{Cornalba:2009ax}.}
\begin{equation}
\mathcal{T}(z)=i\left(1-e^{iQ^{2}Q_{s}^{2}z^{4}}\right),\label{scattampQs}
\end{equation}
and will use only the saturation momentum $Q_{s}$ to refer to the properties of the target. In general, we will be interested in the case where the medium appears dense to the projectile and multiple scatterings have to be considered. Specifically $Q^{2}\ll Q_{s}^{2}$.

In order to get further insight into the scattering piece of the field, let's take a closer look at the Green's function in (\ref{fullsollg}). Considering the Fourier transform in the $x^{-}$ variable
\begin{equation}
G_{L}(z,z';x^{-}-y^{-})=\int\frac{dk^{+}}{2\pi}e^{-ik^{+}(x^{-}-y^{-})}G_{L}(z,z';K^{2})\, ,\label{greenfourier}
\end{equation}
with $K^{2}=-2q^{-}k^{+}$, we can solve Eq. (\ref{greeneq}) in momentum space. The space-like case $K^{2}>0$ is given by \cite{Avsar:2009xf}
\begin{equation}
G_{L}(z,z';K^{2})=-zz'(\BK_{0}(Kz')\BI_{0}(Kz)\Theta(z'-z)+\BK_{0}(Kz)\BI_{0}(Kz')\Theta(z-z'))\, ,
\end{equation}
and the time-like case $K^{2}<0$ by its analytic continuation
\begin{equation}
G_{L}(z,z';K^{2})=-\frac{i\pi zz'}{2}(\BH^{(1)}_{0}(|K|z')\BJ_{0}(|K|z)\Theta(z'-z)+\BH^{(1)}_{0}(|K|z)\BJ_{0}(|K|z')\Theta(z-z'))\, .\label{tlgreenmom}
\end{equation}

Inserting Eq. (\ref{greenfourier}) into \eqref{fullsollg}, we get an explicit expansion of the scattering piece in terms of definite momentum states. For large $z$, the terms with $\Theta(z'-z)$ can be neglected, in which case this expansion is written as a superposition of vacuum states. Moreover, the space-like states can be neglected as well since they fall off exponentially. The resulting expression is much simpler and involves only time-like states.

The relevant scattering part to be used at large $z$ is
\begin{align}
A^{\prime(s)}_{+}(x^{-},z)&\simeq -2q^{-}\int_{0}^{\infty}\frac{dk^{+}}{2\pi}e^{-ik^{+}x^{-}}\int\frac{dz'}{z'}\left(-\frac{i\pi zz'}{2}\BH_{0}^{(1)}(|K|z)\BJ_{0}(|K|z')\right)\mathcal{T}(z')A_{+}^{\prime(0)}(0,z')\, ,\\
&=-\frac{i}{4}q^{-}Q^{2}\tilde{\mathcal{A}}_{L}\int_{0}^{\infty}dk^{+}e^{-ik^{+}x^{-}}\left(\int dz'\; z'\BJ_{0}(|K|z')\mathcal{T}(z')\BK_{0}(Qz')\right)z\BH_{0}^{(1)}(|K|z)\, .\label{exptlmodes}
\end{align}

A similar analysis can be applied to the transverse components. The solution to Eq. (\ref{treom}) can be written as
\begin{equation}
A_{i}(x^{-},z)=A_{i}^{(0)}(x^{-},z)-2q^{-}\int \frac{dz'}{z'}G_{T}(z,z';x^{-})\mathcal{T}(z')A_{i}^{(0)}(0,z')\, ,\label{fullsoltr}
\end{equation}
where the vacuum solution is
\begin{equation}
A^{(0)}_{i}(x^{-},z)=Q\mathcal{A}_{i}(x^{-})z\BK_{1}(Qz)\, .\label{vacsoltr},
\end{equation}
and the transverse Green's function satisfies
\begin{equation}
(z\partial_{z}z^{-1}\partial_{z}+2iq^{-}\partial_{-})G_{T}(z,z';x^{-}-y^{-})=z\delta(z-z')\delta(x^{-}-y^{-})\, .\label{greeneqtr}
\end{equation}
The momentum space Green's function is given by
\begin{equation}
G_{T}(z,z';K^{2})=-zz'(\BK_{1}(Kz')\BI_{1}(Kz)\Theta(z'-z)+\BK_{1}(Kz)\BI_{1}(Kz')\Theta(z-z'))\, ,\label{slgreentr}
\end{equation}
for the space-like case $K^{2}>0$, and
\begin{equation}
G_{T}(z,z';K^{2})=-\frac{i\pi zz'}{2}(\BH^{(1)}_{1}(|K|z')\BJ_{1}(|K|z)\Theta(z'-z)+\BH^{(1)}_{1}(|K|z)\BJ_{1}(|K|z')\Theta(z-z'))\, ,
\end{equation}
for the time-like case $K^{2}<0$.

Following the same argument as in the longitudinal case, the scattering piece we will use for large $z$ is
\begin{equation}
A^{(s)}_{i}(x^{-},z)\simeq \frac{i}{2}q^{-}Q\tilde{\mathcal{A}}_{i}\int_{0}^{\infty}dk^{+}e^{-ik^{+}x^{-}}\left(\int dz'\; z'\BJ_{1}(|K|z')\mathcal{T}(z')\BK_{1}(Qz')\right)z\BH_{1}^{(1)}(|K|z)\, .\label{exptlmodestr}
\end{equation}

\section{Energy-momentum tensor for the gravity wave and relation with the structure functions}

In order to get a physical picture of the production of the different time-like modes lets take a look at the flow of energy in different directions. The easiest way to do this is to calculate specific components of the energy-momentum tensor associated with the gravity wave. Our starting point will be the following relation between the energy-momentum tensor and the action \cite{landau},
\begin{equation}
\frac{1}{2}\sqrt{-g}T_{\mu\nu}=\frac{\delta S}{\delta g^{\mu\nu}}\, .\label{Tmunu}
\end{equation}

\subsection{Energy flow in the fifth dimension}\label{totalflux}
In particular we are interested in the flux of energy propagating down the fifth dimension, since this is the energy carried away by the particles produced in the collision. For that purpose, the relevant component to calculate is $T_{+z}$. From Eq. (\ref{action}) we get
\begin{align}
\frac{1}{2}\sqrt{-g}T_{+z}&=-\frac{N^{2}}{32\pi^{2}R}\sqrt{-g}F_{+\alpha}F_{z\beta}g^{\alpha\beta}\\
&=-\frac{N^{2}}{32\pi^{2}R}\sqrt{-g}\left(F_{+-}F_{z+}g^{-+}+F_{+i}F_{zj}g^{ij}\right).
\end{align}
Taking into account the gauge condition $A_{z}=0$, assuming there is no $x_{\perp}$ dependence, and plugging in the AdS$_{5}$ metric, we can write an explicit expression for the mixed component $T_{+}^{z}$ in terms of the fields,
\begin{align}
\sqrt{-g}T_{+}^{z}&=\sqrt{-g}T_{+z}g^{zz}\\
&=-\frac{N^{2}}{16\pi^{2}z}\left(-(\partial_{+}A_{-}-\partial_{-}A_{+})\partial_{z}A_{+}+\partial_{+}A_{i}\partial_{z}A_{i}\right).
\end{align}
Using the complex representation of the plane wave fields as in Eq. \eqref{planeA}, we calculate the $x^{+}$-averaged energy flow. This is given by 
\begin{align}
\sqrt{-g}T_{+}^{z}&=-\frac{N^{2}}{32\pi^{2}z}\text{Re}\left((iq^{-}A_{-}+\partial_{-}A_{+})\partial_{z}A_{+}^{*}-iq^{-}A_{i}\partial_{z}A_{i}^{*}\right)\label{emtensor1}\\
&=\frac{N^{2}}{32\pi^{2}z}\text{Im}\left(\frac{1}{q^{-}}z\partial_{z}(z^{-1}A'_{+})A_{+}^{\prime*}-q^{-}A_{i}\partial_{z}A_{i}^{*}\right),\label{emtensor}
\end{align}
where we have used one of the equations of motion (Eq. \eqref{leom1}) in the last step.

To calculate the total energy flow associated with the produced particles we would have to integrate the expression above over positive values of $x^{-}$ (where the scattering piece is localized). By going to large $z$ we can safely extend the $x^{-}$ integration to all values since the incoming field is very small in that region. As will be seen in the following, being able to integrate over all values of $x^{-}$, instead of only positive values, greatly simplifies the calculation.

Again, lets focus in the longitudinal components first. If we write the $+$ component of the field, like in Eq. \eqref{exptlmodes}, in the form
\begin{equation}
A'_{+}(x^{-},z)=\int_{0}^{\infty}dk^{+}\; e^{-ik^{+}x^{-}}a(k^{+})z\BH_{0}^{(1)}(|K|z)\, ,\label{fourierA}
\end{equation}
we can easily see that integrating over $x^{-}$ Eq. (\ref{emtensor}) will identify the $k^{+}$ modes from the field and the conjugate field. Taking the large $z$ form of the Hankel functions, the contribution from the longitudinal part to the energy-momentum tensor is
\begin{equation}
\int dx^{-}(\sqrt{-g}T_{+}^{z})_{L}=\frac{N^{2}}{8\pi^{2}q^{-}}\int_{0}^{\infty}dk^{+}\;|a(k^{+})|^{2}\, ,\label{summodes}
\end{equation}
which is independent of $z$ (as it should). Comparing Eqs. (\ref{fourierA}) and (\ref{exptlmodes}) we get the explicit form of $a(k^{+})$ 
\begin{equation}
a(k^{+})=-\frac{i}{4}q^{-}Q^{2}\tilde{\mathcal{A}}_{L}\int dz'\; z'\BJ_{0}(|K|z')\mathcal{T}(z')\BK_{0}(Qz')\, .\label{fcoeff}
\end{equation}
Inserting this into (\ref{summodes}) and changing the momentum integration to a virtuality integration we get
\begin{align}
\int dx^{-}(\sqrt{-g}T_{+}^{z})_{L}=\frac{N^{2}}{128\pi^{2}}Q^{4}|\tilde{\mathcal{A}}_{L}|^{2}\int dz'\,dz''\,d|K|\;&|K|\BJ_{0}(|K|z')\BJ_{0}(|K|z'')\nonumber\\
&\times z'\BK_{0}(Qz')z''\BK_{0}(Qz'')\mathcal{T}(z')\mathcal{T}(z'')\, .
\end{align}
The virtuality integration gives a $\delta$-function identifying $z'$ with $z''$. Finally, we get
\begin{equation}
\int dx^{-}(\sqrt{-g}T_{+}^{z})_{L}=\frac{N^{2}}{128\pi^{2}}Q^{4}|\tilde{\mathcal{A}}_{L}|^{2}\int dz'\;z' \BK_{0}^{2}(Qz')|\mathcal{T}(z')|^{2}\, .
\end{equation}
A similar analysis can be done with the transverse components of the field. The total result for the energy flux is
\begin{equation}
\int dx^{-}\sqrt{-g}T_{+}^{z}=\frac{N^{2}}{32\pi^{2}}Q^{2}\int dz'\; z' |\mathcal{T}(z')|^{2}\left(\frac{Q^{2}}{4}|\tilde{\mathcal{A}}_{L}|^{2}\BK_{0}^{2}(Qz')+(q^{-})^{2}|\tilde{\mathcal{A}}_{i}|^{2}\BK_{1}^{2}(Qz')\right).\label{Tplusz}
\end{equation}

\subsection{Relation with imaginary part of the action and optical theorem}
As first noted in \cite{yoshinotes} for the case of an infinite plasma, the total flux calculated above is proportional to the imaginary part of the complex classical action. To see this, consider the classical action in the AdS$_{5}$ metric \cite{Avsar:2009xf,yoshinotes} (the shockwave part of the metric doesn't appear explicitly since the integrand is evaluated at $z=0$)
\begin{align}
S_{cl}&=\frac{N^{2}}{32\pi^{2}}\int d^{4}x \left.\frac{1}{z}\left(-A^{*}_{+}A'_{-}-A^{*}_{-}A'_{+}+A^{*}_{i}A'_{i}\right)\right|_{z=0}\label{action1}\\
&=\frac{N^{2}}{32\pi^{2}}\int d^{4}x \left.\frac{1}{z}\left(-\frac{1}{iq^{-}}A^{*}_{+}\partial_{-}A'_{+}-A^{*}_{-}A'_{+}+A^{*}_{i}A'_{i}\right)\right|_{z=0}\, ,\label{action2}
\end{align}
where we have made use of the equations of motion to get rid of the term with $A'_{-}$. Here the relation with the energy-momentum tensor is already visible when comparing Eqs. (\ref{action2}) and (\ref{emtensor1}) since the imaginary part of the action should be $z$ independent (see Ref. \cite{Son:2002sd}). Nevertheless, let us work out the explicit form of the imaginary part of the action in terms of the fields found in the previous section. Taking into account that the scattering piece of the solution to the equations of motion is zero at $z=0$, and that the vacuum classical action is real, we see that the only contribution to the imaginary part comes from $-A^{*(0)}_{+}A^{\prime(s)}_{-}-A^{*(0)}_{-}A^{\prime(s)}_{+}+A^{*(0)}_{i}A^{\prime(s)}_{i}$ in Eq. (\ref{action1}).

Lets consider first the term with the transverse components only. The scattering piece to be considered takes the form
\begin{equation}
A_{i}^{(s)}(x^{-},z)=-2q^{-}\int\frac{dz'}{z'}\frac{dk^{+}}{2\pi}e^{-ik^{+}x^{-}}G_{T}(z,z';K^{2})\mathcal{T}(z')A_{i}^{(0)}(0,z')\, .
\end{equation}
The contribution to the classical action we are interested in is
\begin{align}
(S_{cl}-S_{0})_{T}&=\frac{N^{2}}{32\pi^{2}}\int d^{4}x \left.\frac{1}{z}A^{*(0)}_{i}A^{\prime(s)}_{i}\right|_{z=0}\\
&=-\frac{N^{2}q^{-}}{16\pi^{2}}\int d^{4}x\left[\frac{1}{z}e^{iq^{+}x^{-}}\tilde{\mathcal{A}}^{*}_{i}\int\frac{dz'}{z'}\frac{dk^{+}}{2\pi}e^{-ik^{+}x^{-}}\partial_{z}G_{T}(z,z';K^{2})\mathcal{T}(z')A_{i}^{(0)}(0,z')\right]_{z=0}\, .
\end{align}

The $x^{-}$ integration gives a $\delta$-function which combined with the $k^{+}$ integration picks up the $q^{+}$ component in the Green's function in the scattering piece. The remaining integrand is independent of the rest of the coordinates and therefore we only get an extra factor of volume $V=L_{x}L_{y}L_{+}$. We then arrive at
\begin{equation}
(S_{cl}-S_{0})_{T}=-\frac{N^{2}q^{-}V}{16\pi^{2}}\tilde{\mathcal{A}}^{*}_{i}\int\frac{dz'}{z'}\left[\frac{1}{z}\partial_{z}G_{T}(z,z';Q^{2})\right]_{z=0}\mathcal{T}(z')A^{(0)}_{i}(0,z')\, .
\end{equation}
Inserting the vacuum solution (\ref{vacsoltr}) and using Eq. (\ref{slgreentr}) to evaluate
\begin{equation}
\lim_{z\to0}\frac{1}{z}\partial_{z}G_{T}(z,z';Q^{2})=-z'Q\BK_{1}(Qz')\, ,
\end{equation}
we finally find
\begin{equation}
(S_{cl}-S_{0})_{T}=\frac{N^{2}q^{-}Q^{2}V}{16\pi^{2}}|\tilde{\mathcal{A}}_{i}|^{2}\int dz'\; z'\BK_{1}^{2}(Qz')\mathcal{T}(z')\, .
\end{equation}

Similarly, the first two terms of Eq. (\ref{action2}) give the contribution from the longitudinal components. Carrying out the corresponding calculation, we get as the total scattering contribution to the classical action
\begin{equation}
S_{cl}-S_{0}=\frac{N^{2}Q^{2}V}{16\pi^{2}q^{-}}\int dz'\; z'\mathcal{T}(z')\left(\frac{Q^{2}}{4}|\tilde{\mathcal{A}}_{L}|^{2}\BK_{0}^{2}(Qz')+(q^{-})^{2}|\tilde{\mathcal{A}}_{i}|^{2}\BK_{1}^{2}(Qz')\right).\label{claction}
\end{equation}
From the explicit form of the scattering amplitude \eqref{scattampQs} we know it satisfies the following relation
\begin{equation}
|\mathcal{T}(z')|^{2}=2\text{Im}\mathcal{T}(z')\, .
\end{equation}
Therefore, comparing Eqs. (\ref{Tplusz}) and (\ref{claction}),
\begin{equation}
q^{-}\text{Im} S_{cl}=V\int dx^{-}\;\sqrt{-g}T_{+}^{z}\, .
\end{equation}

Presumably, this relation is valid regardless of the specific target since it can be seen as an illustration of the optical theorem in this supergravity context. By following the propagation of the time-like modes in (\ref{Tplusz}) we are performing a sum over final states instead of calculating the forward scattering amplitude. This relation supports our picture of the time-like modes representing real particles produced in the scattering process and therefore making the scattering inelastic, even though the so called scattering amplitude has the form characteristic of elastic interaction, which is a consequence of the series defining the scattering piece of the field being written in terms of diffractive contributions from graviton exchanges. This apparent confusion is due to the fact that the scattering amplitude appears when working in a coordinate basis where the produced particles are not directly visible except for the decoherence introduced by the $z$ dependence of the phase picked up when going through the shockwave (Eq. (\ref{discphase})).

\subsection{Comparison with initial flux in the $x^{-}$ direction}
The flux just calculated is to be compared with the flux of the incoming gravity wave. Since the incoming probe is assumed to be a left-mover coming from negative large values of $x^{-}$, the relevant component of the energy-momentum tensor is $T_{+}^{-}$. From Eq. (\ref{Tmunu}) we get
\begin{align}
\sqrt{-g}T_{+}^{-}&=\sqrt{-g}T_{++}g^{+-}\\
&=\frac{N^{2}}{8\pi^{2}z}\left((A'_{+})^{2}+(\partial_{+}A_{i})^{2}\right).
\end{align}
In complex notation and using \eqref{planeA},
\begin{equation}
\sqrt{-g}T_{+}^{-}=\frac{N^{2}}{16\pi^{2}z}\left(|A'_{+}|^{2}+(q^{-})^{2}|A_{i}|^{2}\right).
\end{equation}
This quantity is to be calculated for the incoming wave and therefore only the vacuum piece should be used. Using Eqs. (\ref{vacsol}) and (\ref{vacsoltr}),
\begin{equation}
\sqrt{-g}T_{+}^{-}=\frac{N^{2}}{16\pi^{2}}Q^{2}z\left(\frac{Q^{2}}{4}|\tilde{\mathcal{A}}_{L}|^{2}\BK_{0}^{2}(Qz)+(q^{-})^{2}|\tilde{\mathcal{A}}_{i}|^{2}\BK_{1}^{2}(Qz)\right).
\end{equation}

To obtain the total energy flux in the $x^{-}$ direction this result should be integrated with respect to $z$. In that case the only difference between this flux and the one calculated before in the positive $z$ direction (Eq. \eqref{Tplusz}) is a factor of $\frac{1}{2}|\mathcal{T}(z)|^{2}$. In the case of interest $Q\ll Q_{s}$, there is a simple interpretation for this relation between the fluxes. For $z\ll 1/\sqrt{QQ_{s}}$ the scattering amplitude $\mathcal{T}(z)$ is very small, which means that the incoming wave doesn't see the shock wave and goes through unscattered. For large values of $z$ the eikonal phase becomes rapidly oscillating and the factor $\frac{1}{2}|\mathcal{T}(z)|^{2}$ can be safely replaced by 1 in the integration over $z$, giving a total scattering where all the incoming energy flows down the fifth dimension spread among the time-like modes.

Here is a subtle distinction between the longitudinal and the transverse components regarding what fraction of the energy goes unscattered. This fraction is concentrated in the small $z$ region where the two contributions (longitudinal and transverse) have very different behaviors, the longitudinal contribution goes like $z\ln^{2}z$ while the transverse contribution goes like $1/z$. Most of the energy carried by the transverse components will continue its propagation in the $x^{-}$ direction, contrary to what happens with the longitudinal components where essentially all the energy impinging on the target goes toward $z=\infty$ after passing through the shockwave.

\section{Virtuality distribution}

Taking as a starting point Eq. \eqref{summodes}, and the equivalent expression for the transverse components, it is interesting to ask which modes carry most of the energy, or, in other words, which virtualities are favored in the particle production from this DIS process. In order to address those issues let's not perform the $k^{+}$ integration and obtain an (approximate) expression for the coefficient function $a(k^{+})$.

The $z'$ integration in Eq. (\ref{fcoeff}) can't be performed explicitly. In the limit where $Q\ll Q_{s}$ we can make a rough estimate by taking the scattering amplitude $\mathcal{T}(z')$ to be one for $z'>1/\sqrt{QQ_{s}}$ and zero otherwise. In that case, the $z'$ integration takes the form
\begin{align}
\int_{\frac{1}{\sqrt{QQ_{s}}}}^{\infty}dz'\; z'\BJ_{0}(|K|z')\BK_{0}(Qz')=&\frac{1}{Q^{2}+|K|^{2}}\left(\sqrt{\frac{Q}{Q_{s}}}\BJ_{0}\left(\frac{|K|}{\sqrt{QQ_{s}}}\right)\BK_{1}\left(\sqrt{\frac{Q}{Q_{s}}}\right)\right.\nonumber \\
&\left.-\frac{|K|}{\sqrt{QQ_{s}}}\BJ_{1}\left(\frac{|K|}{\sqrt{QQ_{s}}}\right)\BK_{0}\left(\sqrt{\frac{Q}{Q_{s}}}\right)\right) \\
\simeq&\frac{1}{Q^{2}+|K|^{2}}\left(\BJ_{0}\left(\frac{|K|}{\sqrt{QQ_{s}}}\right)-\frac{|K|}{\sqrt{QQ_{s}}}\ln\sqrt{\frac{Q_{s}}{Q}}\BJ_{1}\left(\frac{|K|}{\sqrt{QQ_{s}}}\right)\right),
\end{align}
where we have used the small argument form of the K functions.

Plugging this back in Eq. \eqref{summodes},
\begin{align}
\int dx^{-}(\sqrt{-g}T_{+}^{z})_{L}=\frac{N^{2}Q^{4}}{128\pi^{2}}|\tilde{\mathcal{A}}_{L}|^{2}&\int d|K|\; \frac{|K|}{(Q^{2}+|K|^{2})^{2}}\nonumber \\
&\times\left(\BJ_{0}\left(\frac{|K|}{\sqrt{QQ_{s}}}\right)-\frac{|K|}{\sqrt{QQ_{s}}}\ln\sqrt{\frac{Q_{s}}{Q}}\BJ_{1}\left(\frac{|K|}{\sqrt{QQ_{s}}}\right)\right)^{2}\, .
\end{align}
Instead of trying to perform this integration, we will determine which is the dominant region and therefore determine which virtualities play an important role in the scattering process. The first factor in the integrand is large for $|K|\sim Q$ and falls rapidly for large $|K|$. The second factor is of order 1 for small $|K|$ and becomes large only for $|K| \gg \sqrt{QQ_{s}}$. In that region it grows linearly with $|K|$ but can't compensate for the rapid falling of the first factor which goes as $1/|K|^{3}$. Therefore, the integral is dominated by the region with $|K|\sim Q$.

Now lets turn our attention to the transverse components. From Eq. (\ref{exptlmodestr}) we see that the relevant $z'$ integration takes the form
\begin{align}
\int_{\frac{1}{\sqrt{QQ_{s}}}}^{\infty}dz'\; z'\BJ_{1}(|K|z')\BK_{1}(Qz')=&\frac{1}{Q^{2}+|K|^{2}}\left(\sqrt{\frac{Q}{Q_{s}}}\BJ_{1}\left(\frac{|K|}{\sqrt{QQ_{s}}}\right)\BK_{2}\left(\sqrt{\frac{Q}{Q_{s}}}\right)\right.\nonumber \\
&\left.-\frac{|K|}{\sqrt{QQ_{s}}}\BJ_{2}\left(\frac{|K|}{\sqrt{QQ_{s}}}\right)\BK_{1}\left(\sqrt{\frac{Q}{Q_{s}}}\right)\right) \\
\simeq&\frac{1}{Q^{2}+|K|^{2}}\left(2\sqrt{\frac{Q_{s}}{Q}}\BJ_{1}\left(\frac{|K|}{\sqrt{QQ_{s}}}\right)-\frac{|K|}{Q}\BJ_{2}\left(\frac{|K|}{\sqrt{QQ_{s}}}\right)\right),\\
=&\frac{|K|}{Q(Q^{2}+|K|^{2})}\BJ_{0}\left(\frac{|K|}{\sqrt{QQ_{s}}}\right).
\end{align}
When this expression is used to calculate the energy flow in the fifth dimension we get
\begin{align}
\int dx^{-}(\sqrt{-g}T_{+}^{z})_{T}&=\frac{N^{2}(q^{-})^{2}Q^{2}}{32\pi^{2}}|\tilde{\mathcal{A}}_{i}|^{2}\int d|K|\; \frac{|K|^{3}}{(Q^{2}+|K|^{2})^{2}}\BJ_{0}^{2}\left(\frac{|K|}{\sqrt{QQ_{s}}}\right)\\
&\simeq\frac{N^{2}(q^{-})^{2}Q^{2}}{128\pi^{2}}|\tilde{\mathcal{A}}_{i}|^{2}\ln\frac{Q_{s}^{2}}{Q^{2}}\, ,\label{trlog}
\end{align}
where in the last step we used that the integration is dominated by the region $Q<|K|<\sqrt{QQ_{s}}$. Since the integration is logarithmic, the location of the boundaries of the dominant region don't change the parametric dependence of the final result, and therefore the exact location of the cut-off used for the $z'$ integration doesn't play an important role. Unlike the longitudinal case, for the transverse components the produced modes are distributed over a wide range of virtualities determined by the incoming wave and the properties of the target.

The logarithm found in Eq. (\ref{trlog}) is in agreement with the calculations of the structure functions in \cite{Avsar:2009xf,Mueller:2008bt}.

\section{Localization of the energy flow in coordinate space}

In Section \ref{totalflux} we integrated over $x^{-}$ in order to get an expression for the total flux in the fifth dimension. Instead of performing that integration, we can find the region in $x^{-}$ which contributes the most to this total flux and determine if this flux is localized in coordinate space.

Since the $x^{-}$ integration was what allowed us to identify the $k^{+}$ component of the field and the conjugate field, the expansion in terms of definite time-like modes is no longer useful. In order to focus on the $x^{-}$ dependence of the fields we chose to perform the $k^{+}$ integration at the level of the Green's function.

Consider the longitudinal component first. To be able to perform the $k^{+}$ integration first, it is convenient to write the momentum space Green's function in a general way valid for both time-like and space-like momenta. Specifically \cite{Avsar:2009xf},
\begin{equation}
G_{L}(z,z';K^{2})=-\int_{0}^{\infty}d\omega\;\frac{1}{\omega^{2}+K^{2}}z\BJ_{0}(\omega z)z'\BJ_{0}(\omega z')\, .
\end{equation}
The coordinate space Green's function is then given by
\begin{equation}
G_{L}(z,z';x^{-}-y^{-})=-\int d\omega\frac{dk^{+}}{2\pi}e^{-ik^{+}(x^{-}-y^{-})}\frac{1}{\omega^{2}+K^{2}-i\epsilon}z\BJ_{0}(\omega z)z'\BJ_{0}(\omega z')\, ,
\end{equation}
where the $i\epsilon$ prescription is chosen to obtain a Green's function retarded with respect to $x^{-}$. This choice reproduces correctly the initial condition where we assume a pure plane wave coming from negative values of $x^{-}$ and is also consistent with the momentum space Green's function in Eq. (\ref{tlgreenmom}).

Integrating first over $k^{+}$ and then over $\omega$, we get for the longitudinal Green's function
\begin{equation}
G_{L}(z,z';x^{-}-y^{-})=-\frac{1}{2}\frac{\Theta(x^{-}-y^{-})}{x^{-}-y^{-}}zz'\BJ_{0}\left(\tfrac{q^{-}}{x^{-}-y^{-}}zz'\right)\exp\left[\frac{iq^{-}(z^{2}+z^{\prime2})}{2(x^{-}-y^{-})}\right].\label{explgreen}
\end{equation}
Inserting (\ref{explgreen}) in \eqref{fullsollg}, we see that the scattering piece of the $A'_{+}$ field takes the form
\begin{equation}
A^{\prime(s)}_{+}(x^{-},z)=-\frac{q^{-}Q^{2}}{2}\frac{\Theta(x^{-})}{x^{-}}\tilde{\mathcal{A}}_{L}\;ze^{iq^{-}z^{2}/2x^{-}}\int dz'\; z'\BJ_{0}\left(\tfrac{q^{-}}{x^{-}}zz'\right)e^{iq^{-}z^{\prime2}/2x^{-}}\mathcal{T}(z')\BK_{0}(Qz')\, .
\end{equation}
The $z'$ integration is dominated by the region with $z'\sim 1/Q$ (the integrand is small for small $z'$ and falls exponentially for $z'\gg 1/Q$). In this region, the Bessel function is of order one as long as
\begin{equation}
x^{-}\gtrsim\frac{q^{-}}{Q}z\, .\label{xminusz}
\end{equation}

The phase factor $e^{iq^{-}z^{\prime2}/2x^{-}}$ oscillates rapidly for $x^{-}\ll\frac{q^{-}}{Q^{2}}$ in the dominant region of integration and therefore forces the condition
\begin{equation}
x^{-}\gtrsim\frac{q^{-}}{Q^{2}}\, .\label{cohlength}
\end{equation}
This illustrates the coherence length of the initial system where the produced states cannot be resolved. The small $z$ region is not relevant for our analysis since there we are not able to isolate the contributions from the time-like modes (which flow down the fifth dimension) from the space-like modes (which are non-zero only for small $z$). For large $z$ we only have to consider condition (\ref{xminusz}).

Taking into account that the $x^{-}$ integration of the flux in the fifth dimension is convergent (and therefore the flux becomes small for large values of $x^{-}$), Eq. \eqref{xminusz} implies that the important region in $x^{-}$ for large $z$ is
\begin{equation}
x^{-}\sim\frac{q^{-}}{Q}z\, .\label{trajec}
\end{equation}

The results just derived are also applicable to the transverse components.

Via the UV/IR correspondence, Eq. (\ref{trajec}) can be interpreted as a relation between the longitudinal velocity and the transverse velocity of the produced particles (since the coordinate $z$ is dual to the transverse size). This relation is consistent with the results of \cite{Hatta:2008tx} where it is also shown this relation implies the particles are massless.

\acknowledgments
I would like to thank A.H. Mueller for his guidance and illuminating discussions. I am also grateful to  E. Iancu and Y. Hatta for useful comments. This work is supported by the US Department of Energy.

\end{document}